\documentclass[]{iopart}
\begin{document}
\jl{1}
\input psfig
\title[How the geometry makes the criticality]
{How the geometry makes the criticality in two - component
spreading phenomena?}

\author{
N I Lebovka\dag \ddag\
\footnote[3]{To whom correspondence should be addressed.}
\footnote[4]{E-mail: lebovka@roller.ukma.kiev.ua}
and N V Vygornitskii\dag\
}

\address{\dag\ Institute of Biocolloid Chemistry NASU, 2, bulv.
Vernadskogo, Kyiv,252142, Ukraine}

\address{\ddag\ Kyiv Mogyla Academy University, 2, vul. Scovorody,
Kyiv, 252145, Ukraine}

\begin{abstract}
We study numerically a two-component A-B spreading model (SMK model)
for concave and convex radial growth of 2d-geometries. The seed is
chosen to be an occupied circle line, and growth spreads inside the
circle (concave geometry) or outside the circle (convex geometry).
On the basis of generalised diffusion-annihilation equation for 
domain evolution, we derive the mean field relations describing quite
well the results of numerical investigations. We conclude that the
intrinsic universality of the SMK does not depend on the geometry
and the dependence of criticality versus the curvature observed in
numerical experiments is only an apparent effect. We discuss the
dependence of the apparent critical exponent $\chi_{a}$
upon the spreading geometry and initial conditions.
\end{abstract}



\submitted


\section{Introduction}
\label{S1}
The nonequilibrium growth and spreading phenomena are common in
nature. The examples may include such processes as irreversible
adsorption or deposition on the surface, liquid invasion in porous
media, forest fire, crystal growth, evolution of damages in mechanical
or electrical systems, epidemic spreading, etc. Presently, the
different varieties of such phenomena are extensively studied using
the numerical models \cite{r1,r2,r3,r4,r5,r6}. The most simple models
deal with the ideal species of the same size (or type) and the key
attention is paid to investigation of interrelation between the growth
mechanism, surface equilibration efficiency, details of interparticle
interactions and pattern morphology.

But the ideal systems of monospecies quite rarely occur in nature and
practically, certain variations of physical or structural properties
are always present in such systems. The nonideal polydisperse or
multicomponent systems are of a great scientific and practical interest
\cite{Ren01,Tarafdar01}. Recently, the various growth models with two
species or phases competition
\cite{SMK,Vandewalle01,Drossel01,Kotrla01,Kotrla02,Batchelor01}, and
the multicomponent Potts growth model \cite{Vandewalle02,Vandewalle03}
were investigated. The two-component growth model of Saito and
Muller-Krumbhaar \cite{SMK} generalises the Eden model \cite{Eden} for
the case of two different species, $A$ and $B$, competition
(SMK-model). The authors \cite{SMK} observe the criticality of the
2d-model for the case of equal growth rates of two components in a
half-space planar geometry, i.e. for a case when the growth starts
from the linear seed line. This criticality results from the rules of
species attachment at the growing front, which exclude the nucleation
of $A$ species in a neighbourhood of domain consisting only of $B$
species, and vice versa. The model reveals the competitive growth of
separate domains and their coarsening. In the limit case of large time
or height of the front, $h$, the power-law decay is found for $N_{AB}$
domains,

\begin{equation}
N_{AB}\sim h^{-\chi},    \label{re1}
\end{equation}
where $N_{AB}$ is the number of domains and $\chi{=}2/3$ is the
critical exponent for meandering interfaces \cite{Derrida01}. Later
on, Vandewalle and Ausloos \cite{Vandewalle01} have come to a
conclusion that the criticality of SMK model is not universal and
depends on the assumed geometry of two-component propagation. They
have shown that if the growth starts from the central sites of initial
$AB$ configuration, the value of the critical exponent $\chi$ may
differ considerably from that observed for the half-space planar
geometry ($\chi{=}2/3$). It should be noted that the geometry can
effect the fractal dimension of DLA-like pattern in a closed cavity,
as it was observed in \cite{Lebovka01}. Recently, Batchelor et al.
\cite{Batchelor02} have shown, that for the usual 2d radial Eden
model, the growing scaling exponent $\beta$ governing the interface
roughening exceeds considerably the value of $\beta{=}1/3$, which is
typical for the stochastic growth on a planar substrate (in 1+1
dimensions), and only approaches this value in the asymptotic limit of
the very large Eden cluster radius.

The present work studies kinetics of the two-component spreading on
curved surfaces in assumption of SMK-model. The evolution of domains,
their diffusion and annihilation takes place on a two-dimensional
square lattice, and spreading is either confined inside the circle
(concave geometry), or takes place outside the circle (convex
geometry). For the infinitely large radius of the circle,
$R{\to}\infty$, we revive the half-space or planar geometry
\cite{SMK}, for other limit case of $R{\to}0$ we consider the
spreading in a free space \cite{Vandewalle01}. So this modified model
allows us to perform more precise control over the influence of
geometry on criticality. Using the mean field diffusion-annihilation
approach and simulations for SMK-model, we calculate the density
profiles for wall boundaries between $A$ and $B$ domains as a function
of distance from the seed wall $h$ for seed circles of different
radius, $R$. It will be shown that even in the case when criticality
remains unchanged, the apparent value of the critical exponent of
domain wall spatial distribution may vary and it depends on the
geometry and initial distribution of $A$ and $B$ domains on the seed
line.

The remainder of the paper is organised as follows: Section \ref{S2}
presents a brief description of the model and simulation details.
Section \ref{S3} is devoted to our results presentation and discussion.
Section \ref{S31} contains theoretical discussions of the issue 
of universality for the competitive growth model with curved seed line.
Here, we use the mean field approach for description of the domain walls
coalescence. In the section \ref{S32} we present the results of a
numerical simulation, performed using the SMK algorithm of
\cite{SMK} in order to test the theory exploited in section \ref{S31}. 
Section \ref{S4} contains our conclusions.

\section{Model and Simulation Details}
\label{S2}

We have used the simple two-component SMK growth model \cite{SMK} with
the equal growth rates of two components. The simulation takes place
on the square lattice. In this model, same in Eden model, new
particles attach the growing cluster only along its perimeter, however
in the SMK model the probability of filling any empty site of
perimeter either by A, or by B species, is in proportion to their
concentration at the surface of propagation front. The growth starts
from the circular seed line and develops as the growth front
propagation for two different cases:
\begin{itemize}
 \item Inside circular closed cavity. This is the case of concave
geometry, or inside propagation (IP) case.
 \item Outside the circular closed cavity. This is the case of convex
geometry, or outside propagation (OP) case.
\end{itemize}

The maximal size of studied systems was $2000\times2000$ grid
points and inter-grid distance was set as $d=1$. For each case, 100-500
samples were averaged.

The simulation was carried out for different ordered initial
configurations, including case of a maximal density of domain
boundaries with short correlation length $\lambda{_d}=1$, where single
A and B specie alternate, such as:

\begin{equation}
\begin{tabular}
{|l|l|l|l|l|l|l|l|l|l|l|l|l|l|}
\hline
  ...&A&B&A&B&A&B&A&B&A&B&A&B&...\\
\hline
\end{tabular}
,\label{ds}
\end{equation}
and, also, in the case of lower densities of domain boundaries, with
larger correlation length ($\lambda{_d}=3$ for the case
displayed),when there is an alternation of larger A and B species
domains with equal length $\lambda_d$, such as
\begin{equation}
\begin{tabular}
{|l|l|l|l|l|l|l|l|l|l|l|l|l|l|}
\hline
  ...&A&A&A&B&B&B&A&A&A&B&B&B&...\\
\hline
\end{tabular}
.\label{dl}
\end{equation}

If the initial number of domain boundaries is equal to $N_{AB}^{0}$,
then the initial linear density of the domains walls will be equal to
$\rho_0=N_{AB}^{0}/L$, where $L=2\pi R$ is the length of the seed
line. For the case displayed as the seed line (\ref{ds}) we have
$\rho_{0}=\rho_{max}\approx 1$ and
\begin{equation}
N_{AB}^{0}=N_{max}\approx2\pi R,   \label{d2}
\end{equation}
and for the more common case displayed as the seed line (\ref{dl})
we have $\rho_{0}\approx 1/\lambda_{d}$ and
$N_{AB}^{0}\approx2\pi R/\lambda_{d}$.

\begin{figure}[tbp]
\centerline{\psfig{figure=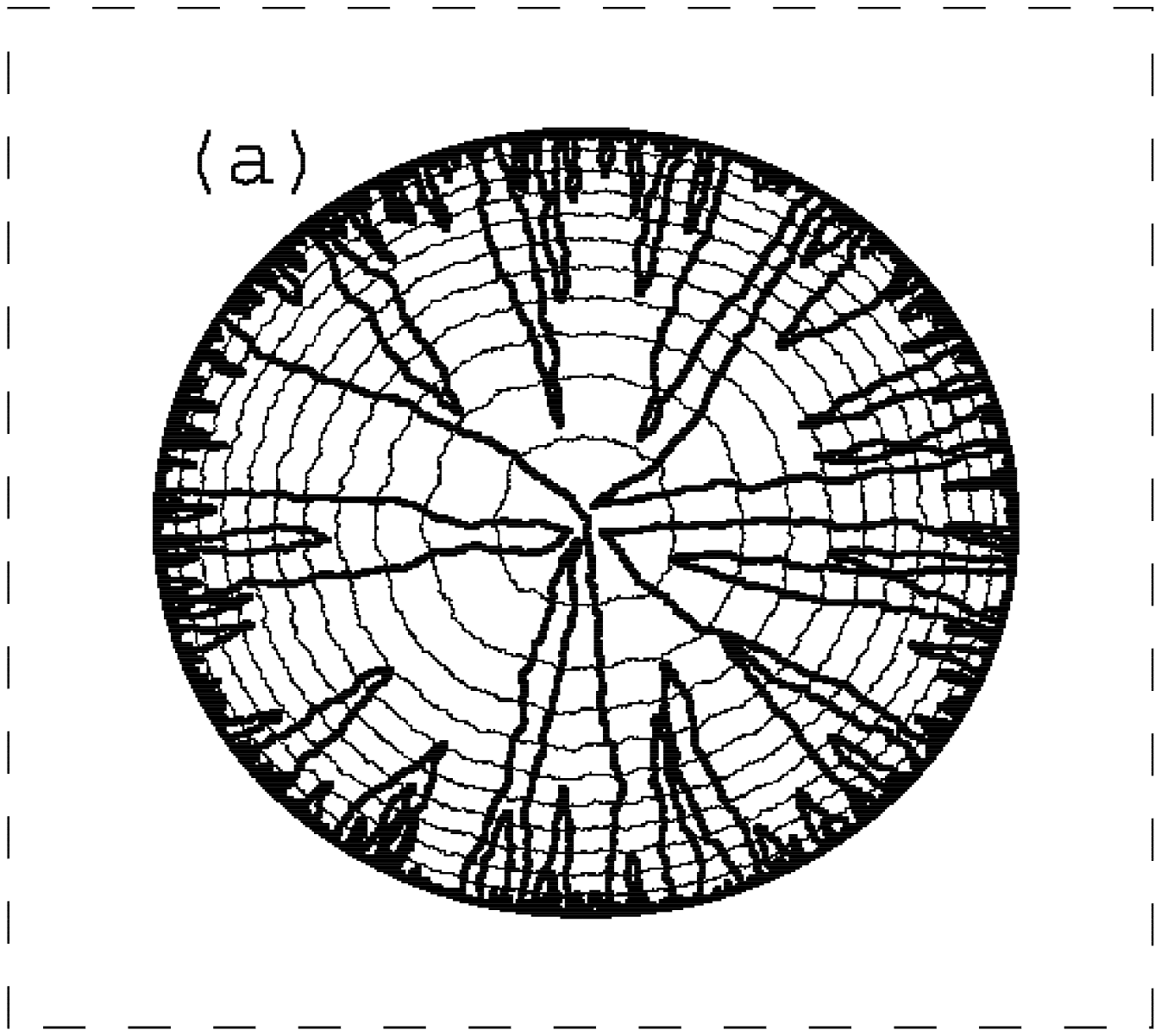,width=8cm,height=8cm,angle=0,clip=}}
\centerline{\psfig{figure=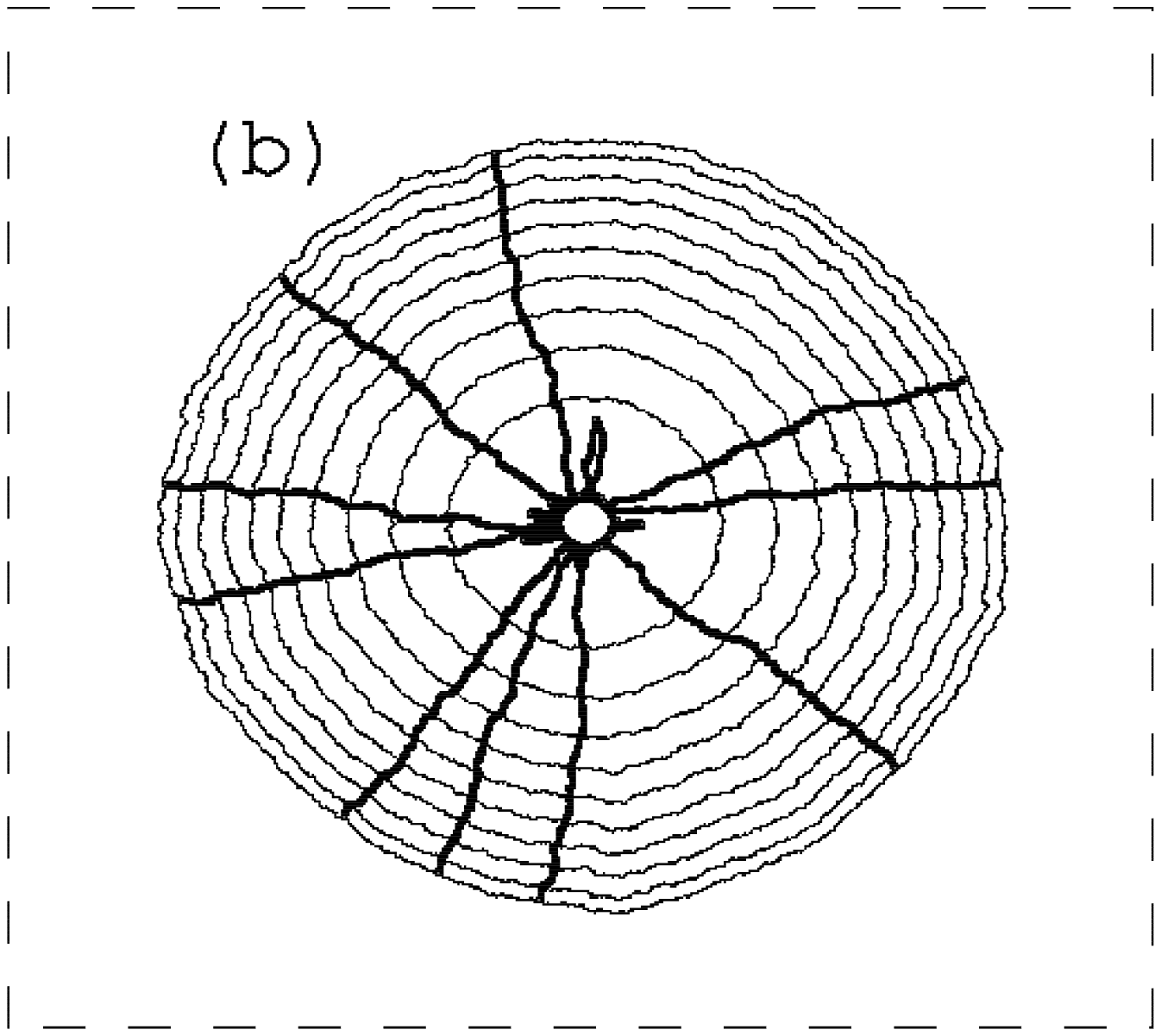,width=8cm,height=8cm,angle=0,clip=}}
\caption{
The typical growth spreading inside the circular cavity of radius
R=1000 (a) and outside the circular exclusion of radius R=60 (b) for
two-species SMK model. The thin concentrical lines are the time
snapshots of fronts for each next group of 300000 particles
attachment. The thick lines show the AB domain interfaces.
}
\label{fig1}
\end{figure}

\Fref{fig1} present the typical spreading pattern of IP case with the
radius of closed cavity R=1000 (a) and the pattern of OP case with
radius of circular exclusion R=60 (b). Here, the thin concentric lines
are the time snapshots of the front evolution as at the moments of
each next $300000$ particles attachment. The thick lines show the $AB$
inter-domain boundaries. In the OP case the simulation terminates when
the maximal radius of growth front $r_{max}$ reaches the boundaries of
system ($r_{max}$=1000). The initial density of boundaries is chosen
to be maximal, $\rho_{0}=\rho_{max}\approx 1$ such as displayed in the
seed line (\ref{ds}). Note, that for both IP and OP cases we observe
the mostly quick decrease of the number of interdomain boundaries
$N_{AB}$ in initial periods of time, near the surface of circular seed
line.

\section{Results and Discussion}
\label{S3}
\subsection{Mean Field  Approach}
\label{S31}

The domain evolution can be described in the term of annihilations
reaction of domain boundary. If we associate each domain boundary with
fantom particle $\widehat{A}$, then we can consider the domain
coarsening as diffusion-annihilation reaction of type
$\widehat{A}+\widehat{A}\to\emptyset$ \cite{SMK}. For the linear seed
line the time evolution of the density $\rho=N_{AB}/L$ satisfies a
mean-field rate equation

\begin{equation}
 \frac{d\rho}{dh}=-a\rho^{1+1/\chi},   \label{e1}
\end{equation}
giving $\rho \sim h^{-\chi}$, where $h$ is the height of the front, or
equivalently, is the time and $a$ is the annihilation rate constant.

We have $\chi{=}1/2$ for the normal Brownian motion
\cite{Smoluchowsky,Bramson,Torney}, and $\chi{=}2/3$ for a two-species
Eden \cite{Derrida01} or SMK \cite{SMK} model.

Using the simulation for the planar geometry, we have estimated the
annihilation rate constant as
\begin{equation}
 a=2.5 \pm 0.2,   \label{e1a}
\end{equation}
where the data used in estimation were averaged over 100 different
samples.

In a case of radial geometry the equation \eref{e1} should be
modified. For this case we have additional source of $\rho$ rate
variations due to the shrinkage or dilation of the spreading front for
the case of growth spreading inside the circle (concave geometry) or
outside it (convex geometry), respectively. The domain density $\rho$
at the distance $r=R\mp h$ from the centre of seed circle is equal to

\begin{equation}
{\rho {=} \frac {N_{AB}} {2\pi (R\mp h)}},                \label{e2}
\end{equation}
where $h$ is the distance from the seed circle, and, here and
hereinafter, the upper sign ($-$) corresponds to the front spreading
inside the seed circle (IP case, concave geometry) and the under sign
($+$) corresponds to the front spreading outside the seed circle (OP
case, convex geometry). Taking the \eref{e2} into account we obtain
instead \eref{e1} the following rate equation for the above-mentioned
cases of radial spreading with $\rho(0){=}\rho_{0}$ as initial
condition:

\begin{equation}
\frac{d\rho}{dh} = -a\rho^{1+1/\chi}-\frac{\rho}{R\mp h}.
\label{e4}
\end{equation}

Introducing the new scaled variables $y{=}\rho R^{\chi}$, and
$x{=}h/R$ we can rewrite \eref{e4} in scaled form
\begin{equation}
   {\frac{dy}{dx} = -a y^{1+1/\chi} - \frac{y}{1 \mp x}}.  \label{e5}
\end{equation}

\Eref{e5} is a Bernoulli equation and 
substitution of $ t{=}y^{-1/\chi} $ reduces it to the differential
equation of the following type
\begin{equation}
 {\frac{dt}{dx} - \frac{t}{1 \mp x} = \frac{a}{\chi}}.    \label{e6}
\end{equation}
The solution of \eref{e6} with initial condition of
$t(0)=t_0={\rho_0^{-1/\chi}}/R$ has the following form
\begin{equation}
{t(x) = t_0 (1 \mp x)^{1/\chi}\left[\pm V(1\mp x)^{1-1/\chi} +1 \mp V\right]},
      \label{e7}
\end{equation}
where
\begin{equation}
V{=}a/(t_{0}(1-\chi))=aR{\rho_0^{1/\chi}}/(1-\chi).
      \label{e71}
\end{equation}

Taking into account the inverse substitution $ y{=}t^{-\chi} $, we
obtain the following final solution of differential equation
\eref{e5}

\begin{equation}
     {y(x)=\frac{y_0}{(1 \mp x)
          \left[\pm V(1 \mp x)^{1-1/\chi}+ 1 \mp V \right]^\chi }},
          \label{e8}
\end{equation}
where $ y_0{=}\rho_0 R^{\chi}$.

For the IP case, when the spreading takes place inside the circle,
the value $y(x)$ goes through the minimum at the point
\begin{equation}
x_{min}(V)=\cases {
1-\left(\frac{\chi}{1-1/V}\right)^{\frac{\chi}{\chi-1}}
&for $V \ge 1/(1-\chi)$,\\
0&for $V < 1/(1-\chi)$.\\}
\label{e9}
\end{equation}

For the case when $V \to \infty$, we have
\begin{equation}
x_{min}= \lim_{V\to\infty}x_{min}(V){=}1-\chi^{\frac{\chi}{\chi-1}},
     \label{e10}
\end{equation}
and 
\begin{equation}
    y_{min}= \lim_{V\to\infty} y(x_{min}){=}
           {a^{-\chi} \chi^\frac{-\chi^2}{1-\chi}}.
     \label{e11}
\end{equation}

\begin{figure}[tbp]
\centerline{\psfig{figure=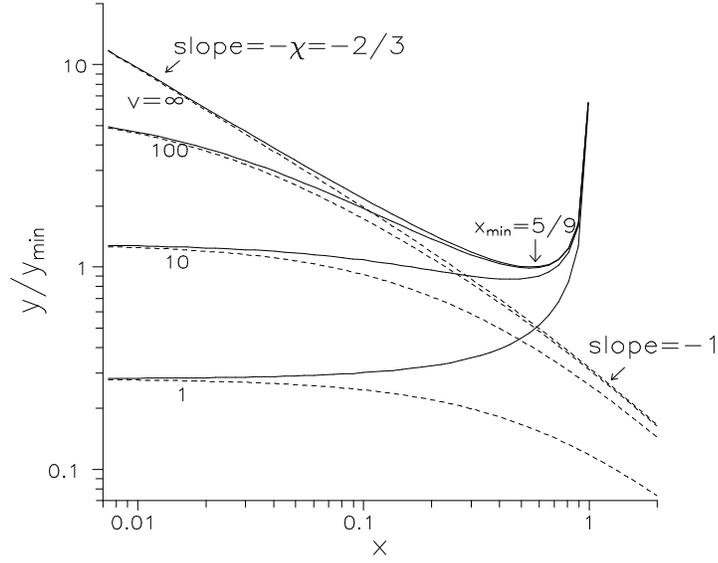,width=10cm,angle=0,clip=}}
\caption{
The mean field dependence of the scaled density $y/y_{min}$ ($y=\rho
R^{\chi} $) versus scaled height of the front $x=h/R$ for the growth
spreading inside the circular cavity (solid lines) and outside the
circular exclusion (dashed lines) at different values of
$V{=}aR{\rho_0^{1/\chi}}/(1-\chi)$. The value of $y_{min}$ as defined
by \eref{e11} corresponds to the minimum of $y(x)$ function in the
point $x=x_{min}=5/9$ (for $\chi=2/3$, see \eref{e10}) for growth
spreading inside the circular cavity in the limit of $V\to\infty$.
}
\label{fig2}
\end{figure}

\Fref{fig2} presents the $y/y_{min}$ versus $x$ dependencies obtained
from \eref{e8},\eref{e11} for $\chi{=}2/3$ at different values of $V$
for the case of growth spreading inside the circle (solid lines) and
outside the circle (dashed lines). The slope value -2/3 accounts for
the slope of lines in the limit $V\to \infty $ for the spreading both
inside and outside the circle, and the slope value -1 accounts for the
slope of the lines in the limit of $x\to \infty $ for the growth
spreading outside the circle.

From \eref{re1}, we can define the apparent value of the critical
exponent $\chi_{a}$ as
\begin{equation}
    {\chi_{a}=-\frac{d\ln N_{AB}}{d\ln h}}. \label{e12}
\end{equation}

\begin{figure}[tbp]
\centerline{\psfig{figure=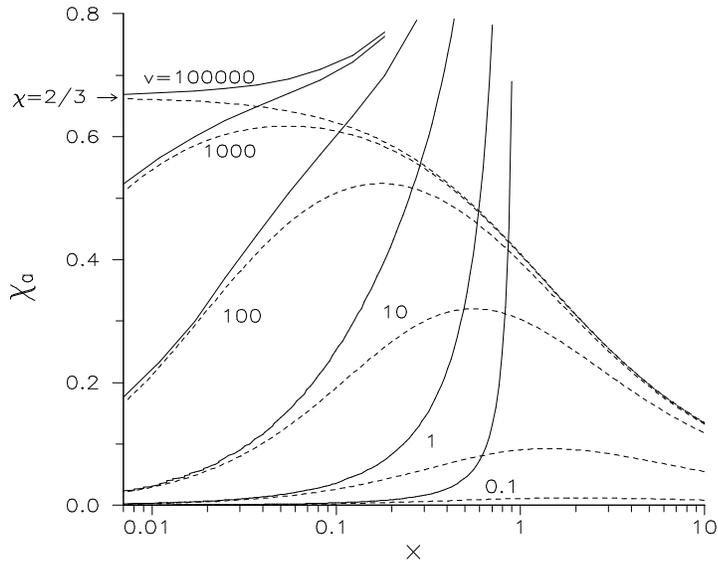,width=10cm,angle=0,clip=}}
\caption{
The effective critical exponent $\chi_{a}$ versus scaled distance from
the seed surface $x=h/R$ for both concave (dash lines) and convex
(solid lines) geometries at different values of
$V{=}aR{\rho_0^{1/\chi}}/(1-\chi)$. Arrow shows the classical value of
$\chi=2/3$ for the planar geometry.
}
\label{fig3}
\end{figure}

\Fref{fig3} shows a plot of the apparent critical exponent $\chi_{a}$
versus scaled distance from the seed surface $x$ for both concave
(dash lines) and convex (solid lines) geometries. We see that for the
propagation started from the curved seed line the value of the apparent
exponent $\chi_{a}$ is not constant and strongly depends on $x$.

Not far from the seed surface, at $x \to 0 $ and high initial density
$\rho_{0}=\rho_{max}\approx 1$ and in the limit $V \to \infty $, the
intrinsic value $\chi_{a}=\chi=2/3 $ is observed. The observed $\chi_{a}$
values substantially decrease with initial density decrease. With
increase of $x$, the value of $\chi _{a} $ always increases for the IP
case. For the OP case, the $\chi _{a} $ goes through the maximum. We
should stress that these are only apparent changes of critical
exponents, estimated on the basis of \eref{e12}, because all
calculations were done for the universal system in assumption of
constancy of intrinsic $\chi$-value ($\chi=2/3$).

\subsection{Numerical Simulation}
\label{S32}

\begin{figure}[tbp]
\centerline{\psfig{figure=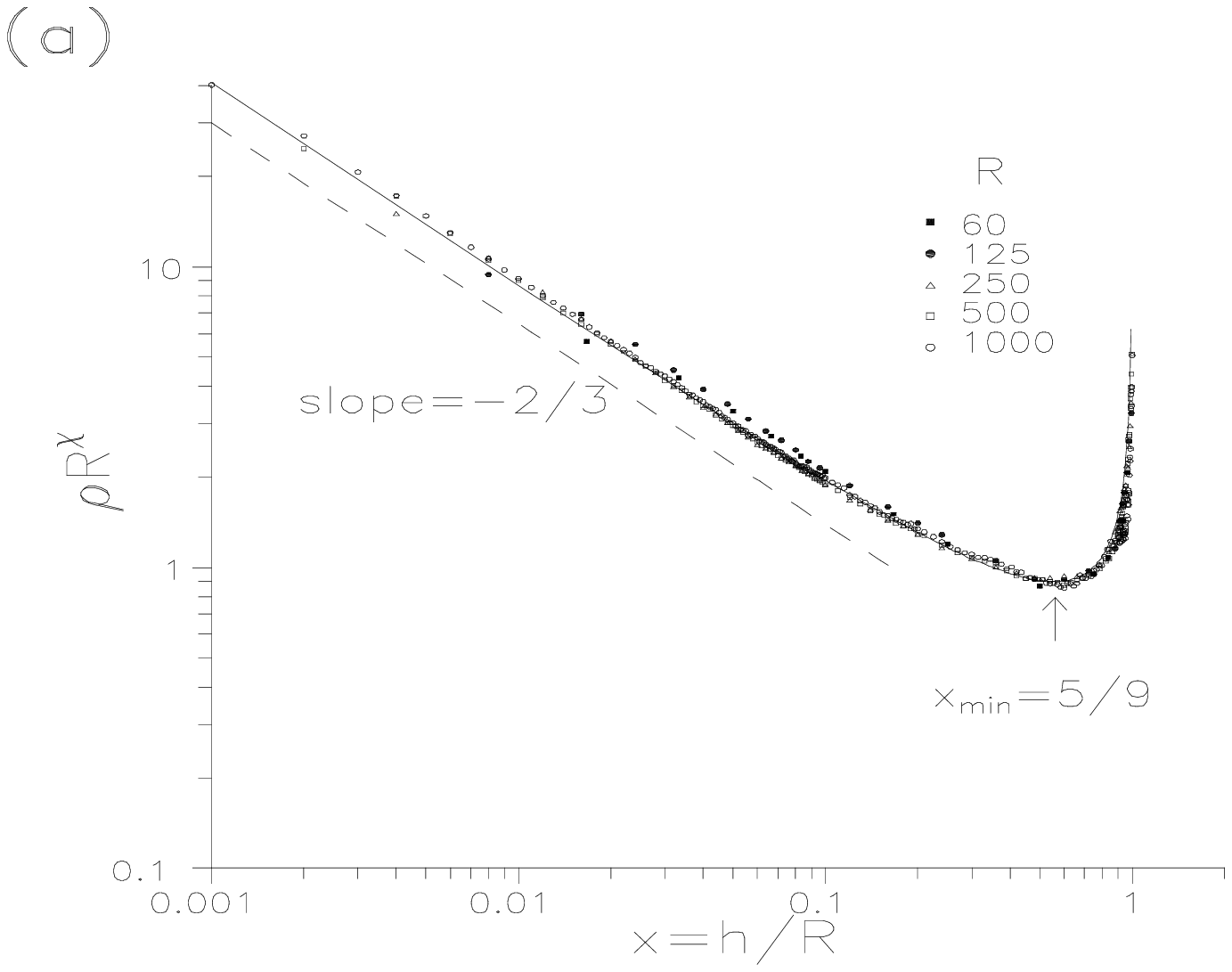,width=12cm,angle=0,clip=}}
\centerline{\psfig{figure=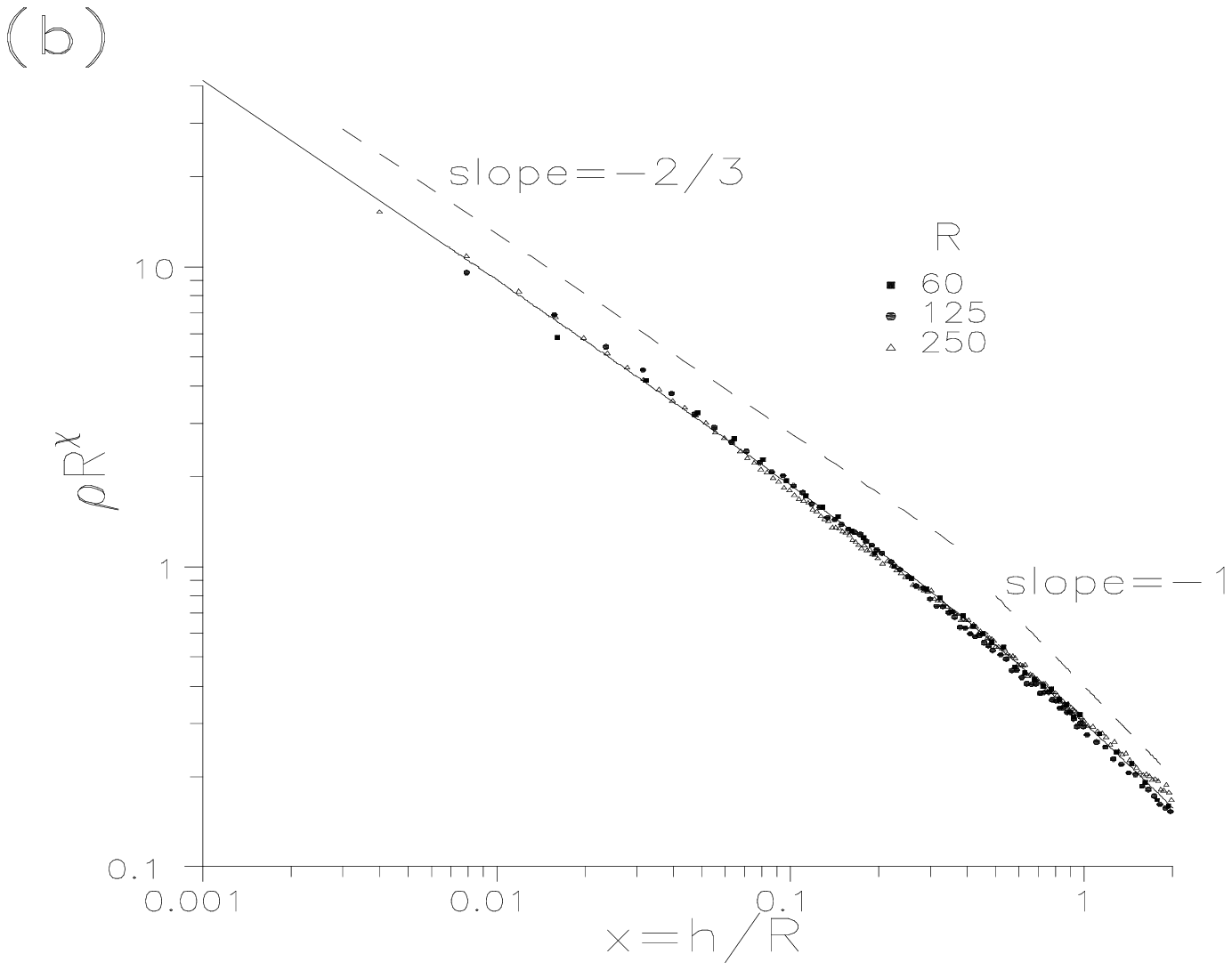,width=12cm,angle=0,clip=}}
\caption{
The scaled density $y=\rho R^{\chi} $ versus scaled height of the
front $x=h/R$ for the growth spreading inside the circular cavity (a)
and outside the circular exclusion (b). The different points are
the simulation results for different cavity (a) or exclusion (b)
radiuses, which are presented on the figures. All data are obtained
for the limit case of $\rho_{0}=\rho_{max}$, for configuration shown
by line (\ref{ds}). The solid lines are obtained from the mean field
equation \eref{e8} where we used $a=2.5 \pm 0.2$ \eref{e1a} estimated
by simulation for the case of planar geometry. Point of minimum
$x=x_{min}=5/9\approx 0.561$ is defined by the \eref{e10} for
$\chi=2/3$.
}
\label{fig4}
\end{figure}

The dependencies of $y{=}\rho R^{\chi}$ upon $x{=}h/R$ obtained as
results of numerical simulations are depicted in \fref{fig4} for the
IP case (a) and OP case (b). The different points correspond to the
different radii of initial circular seed lines. All these data are
obtained for the case of the maximal initial density of domain
boundary $\rho_{0}=\rho_{max}\approx 1$ with ordered configuration of
$A$ and $B$ the species, as in the seed line (\ref{ds}). The continuum
limit for this case corresponds to the situation when $R\gg 1$ and we
can neglect the lattice discreteness. The solid lines in \fref{fig4}
obtained from \eref{e8} for the case of $\chi{=}2/3$, $V\to\infty$ in
the continuum limit. We see that in this limit the coincidence between
the results of the mean field approximation and the computer
simulation is rather good. It is important to note that in the limit
of higher initial density of $AB$ boundaries, all data obtained at
different values of $R$ in the scaled co-ordinates $\rho R^{\chi}$
versus $x{=}h/R$ fall on the same curves for both IP and OP cases.

\begin{figure}[tbp]
\centerline{\psfig{figure=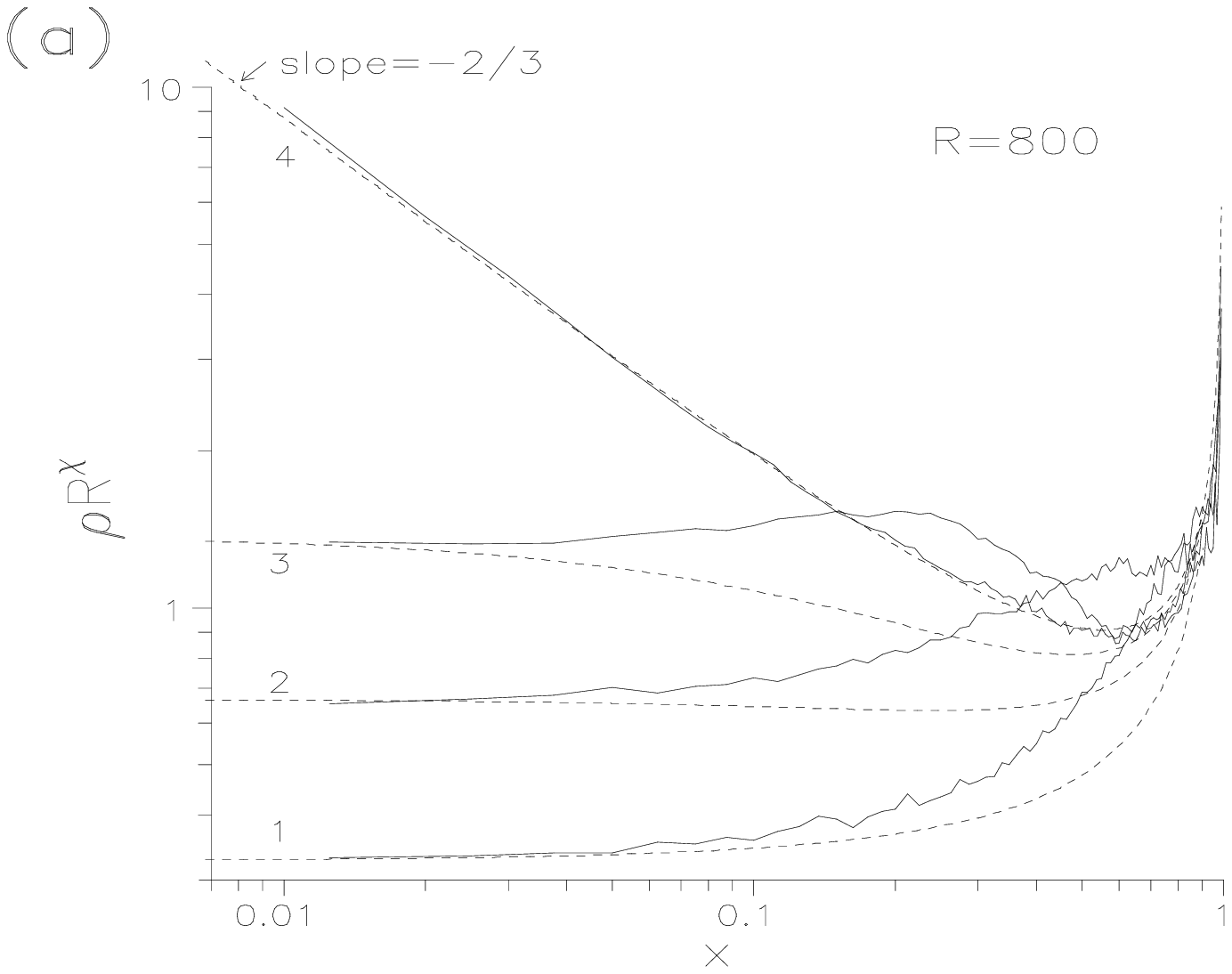,width=12cm,angle=0,clip=}}
\centerline{\psfig{figure=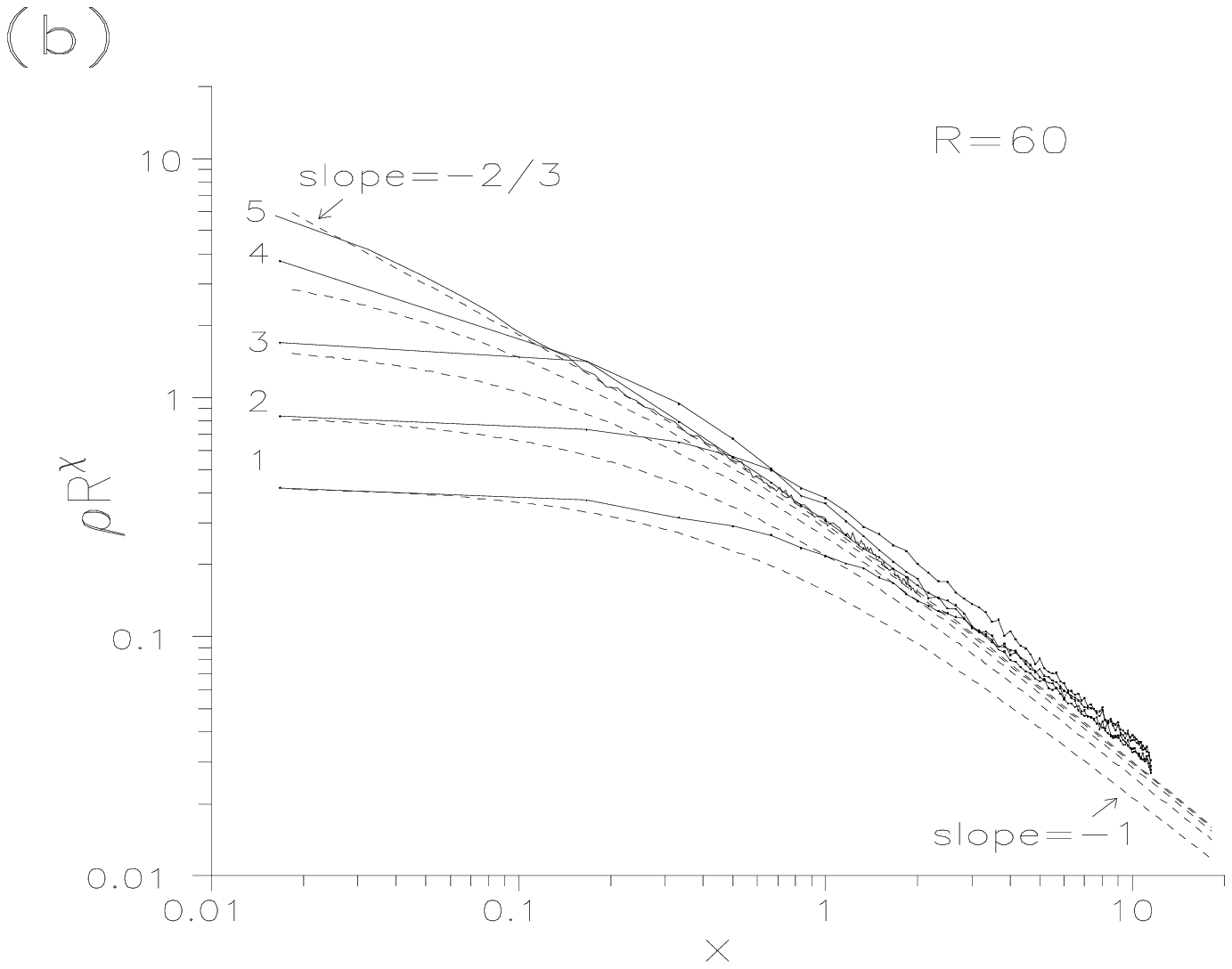,width=12cm,angle=0,clip=}}
\caption{
The scaled density $y=\rho R^{\chi} $ versus scaled height of the
front $x=h/R$ for the growth spreading inside the circular cavity or
radius $R=800$ (a) and outside the circular exclusion or radius
$R=60$ (b).
The solid irregular lines are the simulation results for different
initial numbers of domain interfaces 
(a): $N_{AB}^{0}=20 (1), 40 (2), 80 (3), N_{max} (4)$, and
(b): $N_{AB}^{0}=10 (1), 20 (2), 40 (3), 90 (4), N_{max} (5)$.
Dashed lines are obtained for the same numbers of domain
interfaces through the mean field result \eref{e8} where we used 
$a=2.5 \pm 0.2$ \eref{e1a}, estimated by simulation for
the case of planar geometry.
}
\label{fig5}
\end{figure}

For the relatively small values of initial density $\rho_{0}$, when we
start propagation from configuration with limited number of domain
boundary $N_{AB}\approx 10-100$, the simulation results can deviate
substantially from predictions of the mean-field theory as it is shown
on \fref{fig5} for the IP case (a) and for the OP case (b). The reason
of such deviation is unclear by far but we think that it simply
reflects the limitations of the mean field approach, which follow from
neglecting by the long range correlations between the $AB$ domain
boundaries.

\section{Conclusions}
\label{S4}

We may conclude that the apparent critical exponent $\chi_{a}$ which
is determined through treatment of the numerical experiments results
for SMK model on the basis of \eref{e12}, is not universal or constant
value. For growth spreading in the case of nonplanar geometry the
apparent critical exponent $\chi_{a}$ depends on the height of the
$AB$ front, initial numbers of $AB$ domain and type of the spreading
geometry. This conclusion remains true for SMK model even in the case
when the model preserves its internal universality as defined by the
classical critical exponent $\chi =2/3$. The most good agreement
between results of numerical simulation and of the mean field
approximation is observed only at high initial number of boundaries
$N_{AB}^{0}$. At low initial number of boundaries $N_{AB}^{0}$, the
mean field approximation allows to reach only the qualitatively true
description. In this case, the absence of quantitative description may
be explained by the effect of long-range correlations, which result in
the faults of the mean field description.

We believe, that results obtained in the present work coincide
qualitatively quite well with data presented in \cite{Vandewalle01}.
The authors of this work have shown that for initial configuration
consisting only from two species $AB$ the number of boundaries is
approximately conserved at the level of $N_{AB}\approx 2$ and in this
case they obtained $\chi_{a}\approx 0$. Assuming for propagation under
such conditions $N_{AB}^{0}=2$, $R\approx 1$, $\chi =2/3$ and the
value $a=2.5 \pm 0.2$ \eref{e1a} we have $V \approx 1$ as estimated
from \eref{e71} . As it follows from our data presented at
\fref{fig3}, when the $V$ values are so low, the $\chi_{a}$ value is
small and practically equal to zero at high enough distances from the
seed centre, which is in correlation with data obtained in
\cite{Vandewalle01}. We believe that such apparent alterations of
criticality or scaling exponential functions may occur in other models
of radial growth, as it was observed, e.g., in \cite{Batchelor02} for
Eden model. The similar tasks are of interest for further
investigations with the extended range of models, number of system
components, space dimensionality, etc.

\section*{Acknowledgements}
We are grateful to Marcel Ausloos, Barbara Drossel, Miroslav Kotrla,
Sujata Tarafdar, and Nicolas Vandewalle for providing us with the
preprints of their works and useful correspondence, and we thank
Natalija Pivovarova for help with preparation of the manuscript.
This work is partly supported by a grant QSU082112 from ISSEP.


\newpage

\newpage


\newpage

\newpage

\end{document}